# What could re-infection tell us about $R_0$? a modeling case-study of syphilis transmission

Joshua Feldman[1], Sharmistha Mishra[1,2,3,4]

1. Centre for Urban Health Solutions, St. Michael's Hospital, University of Toronto
2. Department of Medicine, Division of Infectious Disease, University of Toronto
3. Institute of Health Policy, Management and Evaluation, Dalla Lana School of Public Health, University of Toronto
4. Institute of Medical Sciences, University of Toronto

**Corresponding author:** Sharmistha Mishra; sharmistha.mishra@utoronto.ca

**ABSTRACT**

Many infectious diseases can lead to re-infection. We examined the relationship between the prevalence of repeat infection and the basic reproductive number ($R_0$). First we solved a generic, deterministic compartmental model of re-infection to derive an analytic solution for the relationship. We then numerically solved a disease specific model of syphilis transmission that explicitly tracked re-infection. We derived a generic expression that reflects a non-linear and monotonically increasing relationship between proportion re-infection and $R_0$ and which is attenuated by entry/exit rates and recovery (i.e. treatment). Numerical simulations from the syphilis model aligned with the analytic relationship. Re-infection proportions could be used to understand how far regions are from epidemic control, and should be included as a routine indicator in infectious disease surveillance.

**ABBREVIATIONS**

S-I-S susceptible-infectious-susceptible

S-I-R susceptible-infectious-recovered

S-E-I-R-S susceptible-exposed-infectious-recovered-susceptible

# 1.    INTRODUCTION

Infectious disease surveillance systems use observed data to gauge epidemic spread, persistence, and control (Wolicki et al., 2016). Most surveillance relies on per-capita rates of diagnosed cases to characterize disease persistence, with or without the help of disease transmission models (Bansal, Chowell, Simonsen, Vespignani, & Viboud, 2016). Many infectious diseases can be acquired more than once and transmission models that explicitly capture re-infection demonstrate their influence on peak incidence (Barros & Pinho, 2017; Brewer, Peterman, Newman, & Schmitt, 2011; Martinello et al., 2017; Mutsaka-Makuvaza et al., 2018). However, what remains unknown is whether and how observed cases of re-infections could help characterize how far a locale may be from epidemic control (Barros & Pinho, 2017).

Re-supply of susceptible individuals is important in the persistence of infectious disease transmission and the loss of herd immunity. The replenishing of the susceptible population through treatment, and consequent re-infection, is of particular interest in the study of syphilis epidemics. Rates of syphilis continue to rise across the globe and particularly in industrialized settings, where the proportions of new diagnoses that represent re-infection are also increasing (Fenton et al., 2008; Shakeri A, 2017). As a result, syphilis screening and other interventions are now being directed towards those with a history of documented syphilis to prevent re-infection at the individual-level, and incidence at the population-level (Lapple, Wylie, Ross, & Plourde, 2017).

Re-infection is a part of the natural history of many infectious diseases, with or without treatment. For example, infections naturally cleared by the host without conferring protective immunity are often abstracted as "S-I-S" (susceptible-infectious-susceptible) systems leading to cycles of infection and re-infection (Anderson & May, 1991). When infections confer a protective immunity that is aborted by treatment – we shift systems from "S-I-R" (susceptible-infectious-recovered) to "S-I-R-S" (Anderson & May, 1991).

Syphilis is a sexually transmitted infection whose natural history without treatment follows an "S-E-I-R-S" system, with a short exposed (E) period following inoculation and before infectiousness. Treatment aborts the transition from infectiousness during early syphilis to a 'recovered' or clinically latent phase, wherein the pathogen (*T.pallidum*) remains in the body but cannot be transmitted to a sexual partner. Debate surrounds the duration of waning immunity following treatment in the "recovered" or latent phase of syphilis. While many epidemic models of syphilis use an average of 1 to 10 years of a waning or partial immunity following treatment of the latent stage (Gray et al., 2010; Pourbohloul, Rekart, & Brunham, 2003; Tuite & Fisman, 2016), empirical data on how fast protective immunity is lost remain scarce (Magnuson et al., 1956). What is known, however, is that clinical cases of re-infection have been documented following treatment at all stages of syphilis (Ogilvie et al., 2009; Shakeri A, 2017).

Models of syphilis interventions suggest that a prioritized approach to those with a history of treated infection could be more efficient at reducing transmission, compared with applying the same intervention to all susceptible persons (Tuite A, 2017). These findings hint at re-infection as a marker of epidemic emergence and persistence – and raise questions about whether the proportion re-infection could tell us about how far we are from epidemic control. We therefore sought to determine if the proportion re-infection could be used to characterize the underlying transmission dynamics, specifically the basic reproductive number $R_0$ of epidemics with re-infection like that of syphilis.

## 2.    METHODS

Our investigation into the relationship between proportion re-infection and $R_0$ included two components. We conducted a theoretical analysis of a generic 'S-I-S' system with re-infection followed by numerical simulations of a more complex model to capture the natural history of untreated and treated syphilis. Both models were compartmental deterministic systems of coupled differential equations.

### 2.1 Generic re-infection model

We expanded the standard 'S-I-S' model by separating initial and repeat infections into four states: $S_1$, $I_1$, $S_2$, and $I_2$ (Figure 1a). The open population was homogeneous and remained stable with balanced births and deaths. Individuals entered the population (for example, at the onset of sexual activity) as $S_1$, move into compartment $I_1$ during their first infection, and then recover (for example via treatment) into $S_2$ - and then remain within the '$S_2$-$I_2$-$S_2$' cycle following re-infection ($I_2$) and treatment.

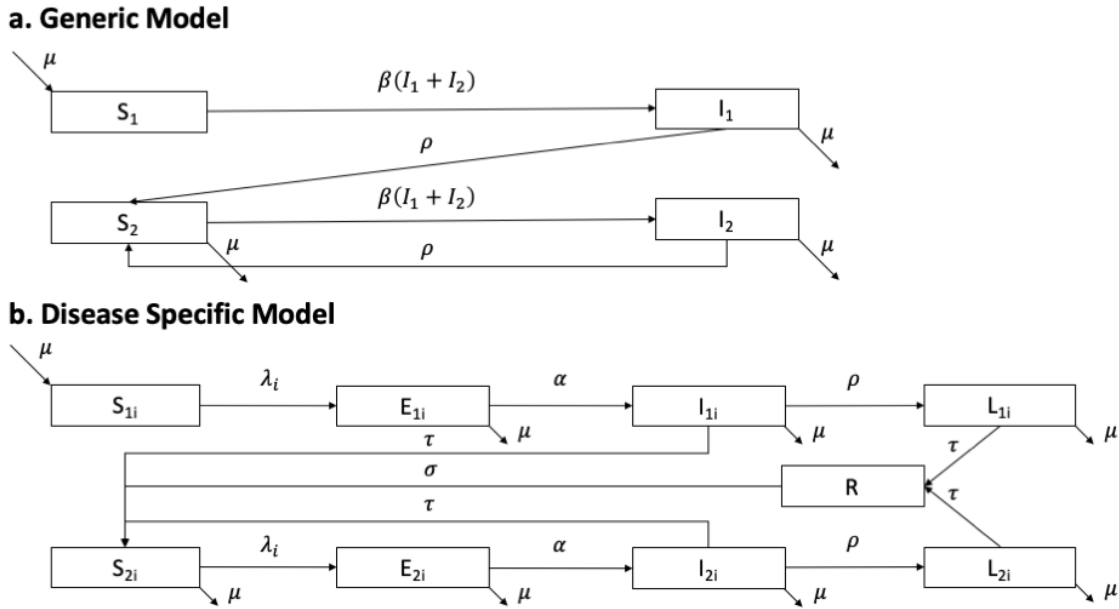

**Figure 1: Model schematic for the (a) generic and (b) disease specific models of re-infection.** Each compartment represents a health state. The parameters describing the rate of transition between compartments of the disease specific model (b) are defined in Table 1. Force of infection $\lambda_i$ is defined in Appendix. Parameters in the generic model are used only for theoretical analysis and are not assigned specific numerical values. The disease specific model reflects a simplified biology of syphilis infection and re-infection.

**Table 1. Parameters for the disease specific model of syphilis.**

| Parameter | Symbol | Unit | Default value or range used in sampling | Source |
|---|---|---|---|---|
| **Population characteristics** | | | | |
| Population Size | $N$ | | 366 279 | Combined from adult ( aged 15-64) male population in Canada(Statistics Canada, 2018b) (12 209 300) and estimated proportion of adult males who report sex with other men (0.03)(Statistics Canada, 2015) |
| Rate of entry into the modeled population | $\mu N$ | Person/Year | 5 831 | Population growth based on birth rate of males(Statistics Canada, 2018a) applied to proportion who have sex with other men |
| **Syphilis natural history** | | | | |
| Transmission probability per sex act | $\beta$ | | Uniform (0.09, 0.64) | (Alexander & Schoch, 1949; Garnett, Aral, Hoyle, Cates, & Anderson, 1997; Moore, Price, Knox, & Elgin, 1963) |
| Incubation period | $1/\alpha$ | Days | 25 | (Garnett et al., 1997) |
| Infectious period | $1/\rho$ | Days | 154 | (Garnett et al., 1997) |
| Protective immunity period | $1/\sigma$ | Years | 5 | (Garnett et al., 1997) |
| **Public health characteristics** | | | | |
| Per-capita treatment rate (per year) | $\tau$ | | Uniform (0.1, 0.8) | (Remis RS et al., 2010 ) |
| **Partnership characteristics** | | | | |
| Average number of partnerships per year (per-capita partner change rate) | $c_{avg}$ | Person | 15 | Generated from survey data - Lambda survey: M-Track Ontario second generation surveillance (2010)(Remis RS et al., 2010 ). In the model (Appendix), $c_H$ refers to partner change rate in the high-activity group, and $c_L$ in the low-activity group. The weighted average results in the default value of $c_{avg}$. |

| Proportion of individuals in the high activity group | $N_H/N$ | | Uniform (0.079, 0.217) | Lower bound: Lambda, Upper bound: M-Track Ontario second generation surveillance (2010)(Remis RS et al., 2010 ) |
|---|---|---|---|---|
| Ratio of high to low activity partnership numbers | $c_H/c_L$ | | Uniform (5, 10) | Assumption |

The differential equations are as follows:

$$\frac{dS_1}{dt} = \mu N - \beta(I_1 + I_2)S_1 - \mu S_1 \qquad (1)$$

$$\frac{dI_1}{dt} = \beta(I_1 + I_2)S_1 - \rho I_1 - \mu I_1 \qquad (2)$$

$$\frac{dS_2}{dt} = \rho(I_1 + I_2) - \beta(I_1 + I_2)S_2 - \mu S_2 \qquad (3)$$

$$\frac{dI_2}{dt} = \beta(I_1 + I_2)S_2 - \rho I_2 - \mu I_2 \qquad (4)$$

where $\mu$ is the entry and exit rate, N is the population, $\beta$ represents the probability of transmission, and $\rho$ is the per capita rate of treatment applied at the same value to the first ($I_1$) or a repeat infection ($I_2$).

The Appendix details the derivation of an explicit relationship between $R_0$ and the prevalence of re-infection, defined as $I_2/(I_1+I_2)$. We set the right-hand-side of the differential equations to 0 to identify the disease-free and endemic equilibria. Following van den Driessche and Watmough, we derived an expression for $R_0$.(Pauline van den Driessche, 2017; P. van den Driessche & Watmough, 2002) Then by expressing the endemic equilibrium in terms of $R_0$, we derived the relationship between $R_0$ and the prevalence of re-infection. Finally, we studied the local stability of this fixed point.

## 2.2 Disease specific syphilis model

To complement our theoretical analysis, we performed numerical simulations of the sexual transmission of syphilis to illustrate the range of observed epidemics among gay, bisexual, and other men who have sex with men (MSM) across Canada (Choudhri Y, Miller J, Sandhu J, Leon A, & J, 2018; Public Health Agency of Canada, 2017). The model is detailed in the Appendix.

We included the following compartments to represent a simplified version of the natural history of syphilis (Figure 1b, Table 1):(Garnett et al., 1997) susceptible ($S_1$, $S_2$), exposed ($E_1$, $E_2$), infectious syphilis ($I_1$, $I_2$), and latent (non-infectious) syphilis ($L_1$, $L_2$). As with the generic re-infection model, we stratified the system into first time and repeat infections. Based on the current understanding of the natural history following treatment, we assumed that individuals treated in the $I_1$ compartment became immediately susceptible to future infections (i.e. moved into $S_2$). However, treatment while in the latent stage meant that individuals experienced a period of waning immunity before becoming susceptible to re-infection.

We included heterogeneity in risk via two sexual activity-levels ($c_H$ [high-activity], and $c_L$ [low-activity]), and assumed proportionate mixing when partnerships were formed (Garnett & Anderson, 1994). The population size stayed constant and individuals remained within their activity groups.

To examine the relationship between proportion re-infection and $R_0$, we generated 4900 synthetic epidemics at equilibrium within the observed rates of annual diagnoses of infectious syphilis among MSM in urban centers across Canada (<1100 cases per 100 000 MSM) (Choudhri Y et al., 2018; Public Health Agency of Canada, 2017). We used biological and sexual behaviour data from the literature to parameterize the model (Table 1). We drew parameters from surveys of the MSM community across Canada, Canadian demographic data, and reviews of the biological properties of syphilis. To generate the synthetic epidemics, we used latin hypercube sampling across the following parameters assuming uniform distributions: biological probability of transmission per partnership ($\beta$); proportion of individuals in the high activity group ($N_H/N$), the ratio of partner change rate among the high and low activity groups ($c_H/c_L$); and the treatment rate ($\tau$). To derive $R_0$, we calculated the next generation matrix to find the dominant eigenvalue (Appendix) (Pauline van den Driessche, 2017; P. van den Driessche & Watmough, 2002).

## 3. RESULTS

### 3.1 Generic model

From our analysis of the next generation matrix, $R_0$ for the generic model is equivalent to the standard SIS model with $R_0 = \frac{\beta}{\rho + \mu}$. Using the fact that a standard SIS model has an endemic equilibrium at $(S^*, I^*) = \left(\frac{1}{R_0}, 1 - \frac{1}{R_0}\right)$, we found that our general model has unique disease-free and endemic equilibria at

$$(S_1^{**}, I_1^{**}, S_2^{**}, I_2^{**}) = (N, 0, 0, 0) \text{ and } (S_1^*, I_1^*, S_2^*, I_2^*) = (S^*(1-\kappa), I^*(1-\kappa), S^*\kappa, I^*\kappa) \quad (4)$$

and where,

$$\kappa = \frac{R_0 + 1}{R_0 + 1 + \frac{\mu}{\rho} R_0}. \quad (5)$$

This means that at the endemic equilibrium, the proportion re-infection is

$$\frac{I_2^*}{I^*} = \kappa = \frac{R_0 + 1}{R_0 + 1 + \frac{\mu}{\rho} R_0}. \quad (6)$$

By analyzing the local stability of the two equilibria, we find that when $R_0 > 1$, the disease free fixed point is unstable and the endemic fixed point is stable. The opposite is the case when $R_0 < 1$.

Equation (6) indicates that $R_0$ increases with proportion re-infection via a non-linear monotonic relationship. Each incremental increase in the proportion re-infection indicates a larger and larger increase in $R_0$. The non-linear relationship is illustrated in Figure 2 under varying treatment rates ($\rho$) and assuming fixed exit rate ($\mu$). The relationship is modified by the ratio of treatment and exit rates: a lower proportion re-infection may signal a large $R_0$ when the baseline treatment rates are low (Figure 2) or when exit rates (such as turn-over in the high-activity population) are high.

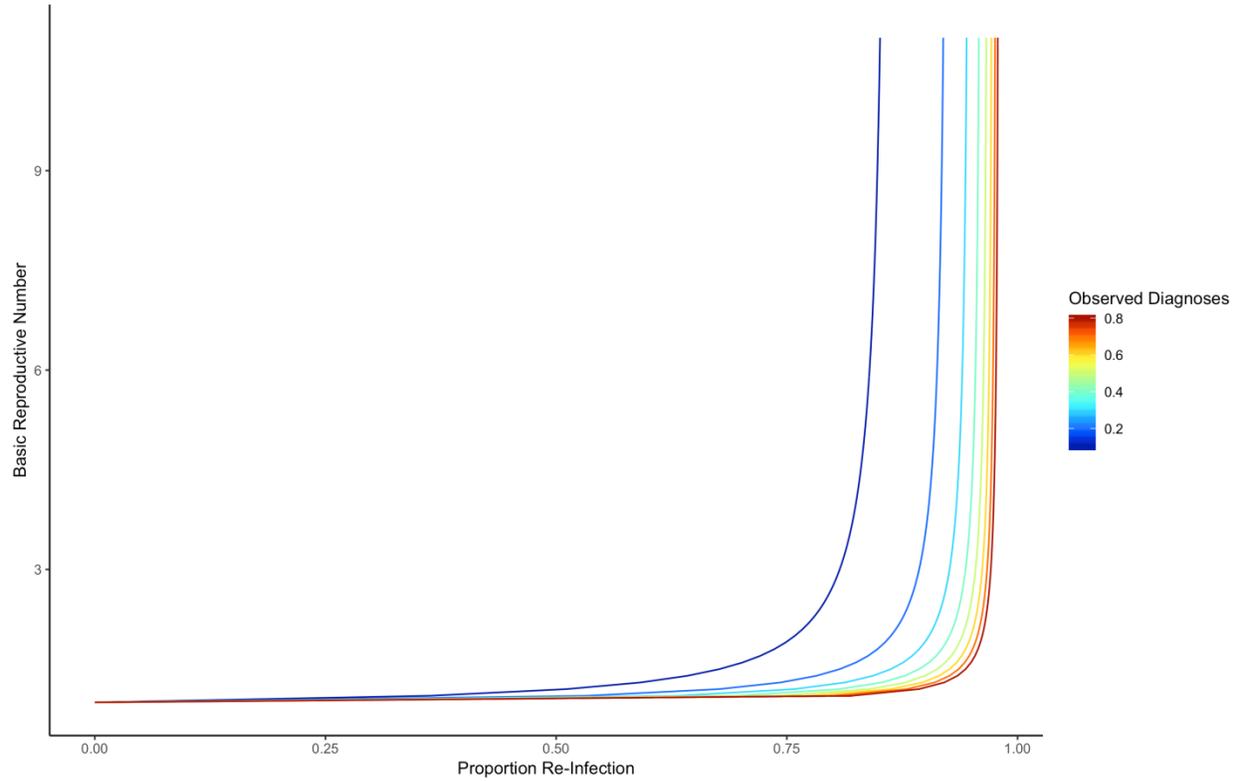

**Fig. 2: Analytic relationship between proportion re-infection and $R_0$, stratified by treatment rate.** Results are based on the generic model.

### 3.2 Disease specific syphilis model

The $R_0$ of the syphilis model is given by (Appendix):

$$R_0 = \frac{\alpha\beta(c_H\ g_H + c_L\ g_L)}{(\alpha+\mu)(\rho+\tau+\mu)} \tag{7}$$

Figure 3 depicts $R_0$ for each modeled proportion re-infection from the synthetic epidemics stratified by the annual diagnoses of syphilis. The non-linear convex relationship between the proportion re-infection and $R_0$ is similar to that derived from the generic model (Figure 2).

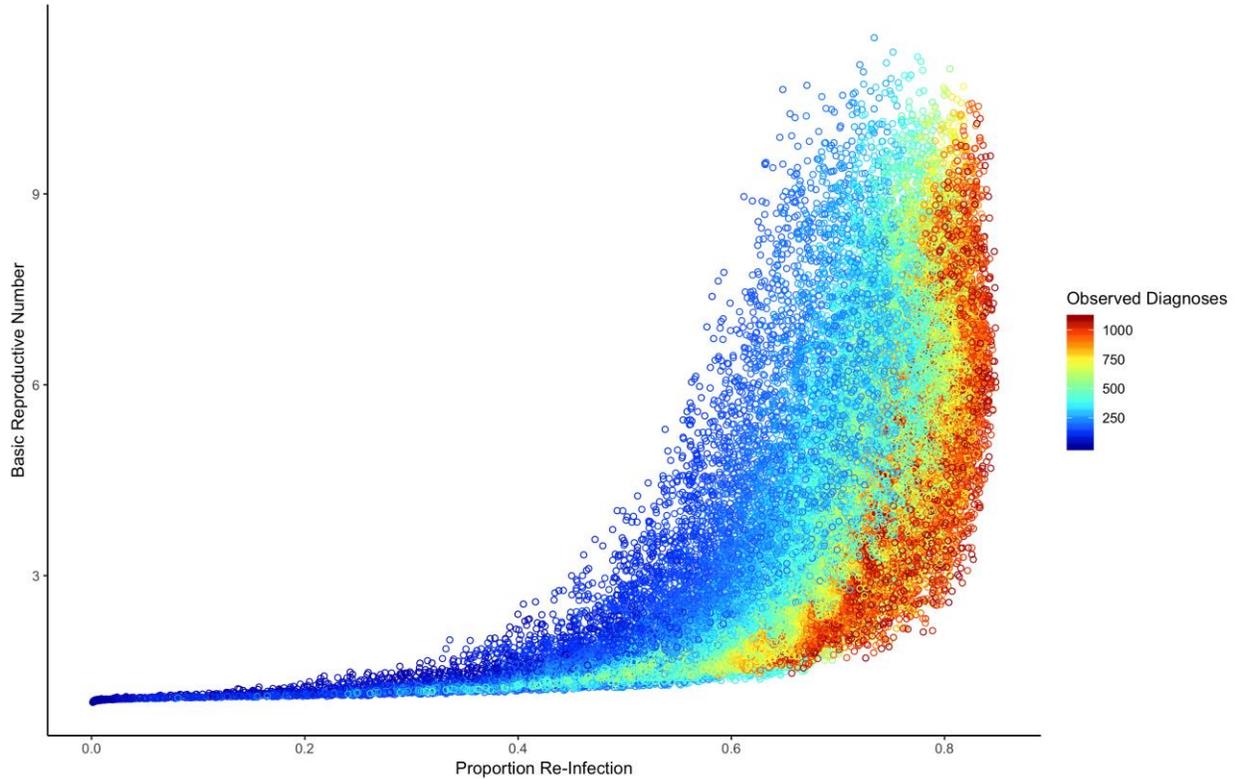

**Fig. 3: Simulations of the relationship between proportion re-infection and $R_0$ by rate of diagnoses.** Results are based on the disease-specific model. Diagnoses rates reflect cases per 100,000 individuals.

For a given $R_0$, higher proportion re-infection and larger annual diagnoses cluster together (Figure 3). Figure 3 also demonstrates that for a given proportion re-infection (for example, 60%), $R_0$ is larger with smaller "observed" epidemics (that is, epidemics with fewer annual diagnoses per 100,000 MSM). The reason for this is depicted in Figure 4. Most of the smaller "observed" epidemics (annual diagnoses, 400-600 per 100,000 MSM) are constrained to very low treatment rates. In contrast, most of the larger "observed" epidemics (annual diagnoses, 800-1100 per 100,000 MSM) are constrained to medium/high treatment rates. The treatment rate also effects how sharply $R_0$ increases with proportion re-infection (Figure 4). Low treatment rates are associated with a steeper increase in $R_0$ per increase in proportion re-infection.

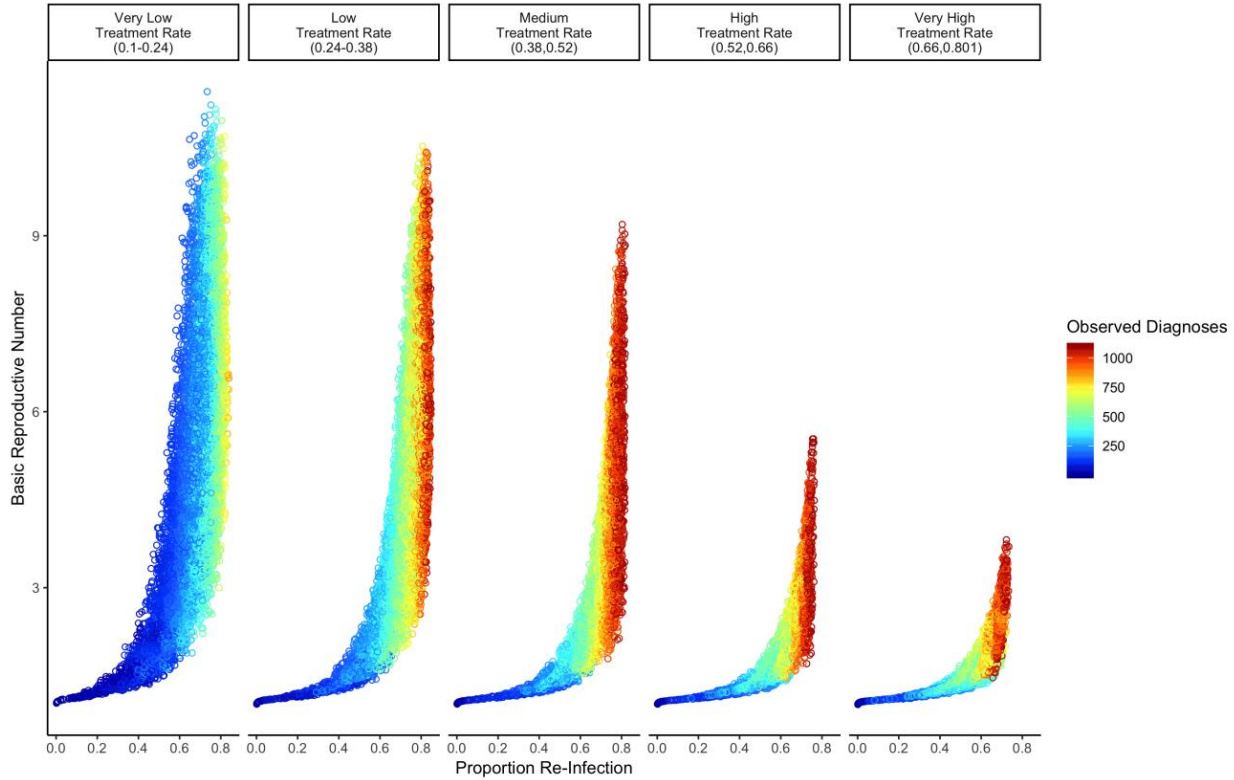

**Fig. 4: Simulations of the relationship between proportion re-infection and R0 stratified by treatment rates.** Results are based on the disease-specific model. Diagnoses rates reflect cases per 100,000 individuals.

## 4.    DISCUSSION

These results have two key implications for surveillance of infectious diseases where repeat infections are possible. First, the proportion re-infection can be used to infer how far a population is from epidemic control in a given setting. Settings with smaller proportion re-infection would be closer to epidemic control, and vice versa.  Due to the non-linearity of the relationship between these two quantities, the effect of reductions or increases in the proportion re-infection on $R_0$ is particularly dramatic when the proportion re-infection is high. For example, if we assume similar rates of syphilis treatment in two Canadian cities – the epidemic is harder to control in the city with the higher proportion re-infection. Although we did not perform any counterfactuals, this relationship would be consistent with other studies that suggest prioritizing interventions for those infected at least once could be an efficient control strategy (Tuite A, 2017).

The second implication for disease surveillance is the unexpected relationship between rates of observed cases and $R_0$. Our synthetic epidemics clustered such that for a given proportion re-infection, lower rates of diagnosed syphilis were associated with a higher $R_0$. Variations in treatment explain this counterintuitive finding. Since the rate of new diagnoses is a function of both the true incidence and the treatment rate, a low observed incidence could stem from either low incidence or low case detection rates. The epidemics with few observed cases and a high $R_0$ also are those with a low treatment rate, which means many infections remain undiagnosed. The result is that observed cases are a large underrepresentation of true incidence. Conversely, when treatment rates are high, the observed diagnoses rates more accurately reflect the true incidence, leading to a positive correlation between observed cases and $R_0$. Increased treatment also reduces $R_0$ by shortening the duration of infectiousness. Taken together, this means that observed diagnosis rates, due to a dependence on both the true incidence and the treatment rate, is an unreliable proxy for $R_0$ and should not be used alone to assess how far a given setting is from epidemic control. While the difference between observed versus true incidence is well-established and its implications discussed (Rothman, Greenland, & Lash, 2008), especially in the context of increased testing over time (Miller, 2008), the use of observed cases to directly or inadvertently infer epidemic potential in the public health literature remains (National Collaborating Centre for Infectious Diseases, 2017; Public Health Agency of Canada, 2017). Our findings provide further evidence that when comparing observed per-capita syphilis rates between two cities, the city with the seemingly larger epidemic may not be the one with the more difficult to control epidemic.

Under the assumption that the treatment rate is the same for new and repeat infections, measuring proportion re-infection bypasses the challenges associated with using observed diagnoses to infer $R_0$. Treatment rates affect the numerator and denominator of the proportion re-infection equally, meaning the observed proportion re-infection is equivalent to the true proportion re-infection given our assumptions. Analyzing the proportion re-infection could give a more accurate picture of the epidemic than the rate of new diagnoses, leading to a better understanding of $R_0$. In fact, using the proportion re-infection and the observed diagnoses together gives the best estimate of $R_0$. Our analytical relationship between $R_0$ and the proportion re-infection is stratified by the treatment rate. Since the observed diagnoses and the proportion re-infection also uniquely determine the treatment rate, stratifying our results by the observed diagnoses could thus narrow the inference on $R_0$.

Interpretation of the findings is limited to the analytic and illustrative case study of syphilis transmission. The findings surrounding inference using proportion re-infection do not account for stochasticity - which are particularly important when considering smaller population size (Anderson & May, 1991); nor plausible differences in testing and treatment rates among those with a prior history of infection versus those without (Workowski & Berman, 2006). We also did not study the implications on trends in proportion re-infection over time and its relation to the effective reproductive number. Although findings can be expected to be similar given the relationship between the basic and effective reproductive numbers (Anderson & May, 1991; Pauline van den Driessche, 2017), future work involves examining the utility of monitoring proportion re-infection alongside trends in observed cases to evaluate the impact of control strategies over time.

## 5. CONCLUSION

Using generic and disease-specific mathematical models that captured re-infection, we identified a relationship between proportion re-infection and $R_0$. We derived a generic expression that reflects a non-linear and monotonically increasing relationship between proportion re-infection and $R_0$ and which is attenuated by entry/exit rates and recovery (i.e. treatment). Our findings highlight the potential for systematically reporting and using data on re-infection alongside observed diagnoses as part of disease surveillance for any pathogen that can cause repeat infections. In doing so, public health teams could gain a better understanding of the underlying disease dynamics, allowing them to identify settings and populations where more resources may be required to achieve local epidemic control.

**ACKNOWLEDGEMENTS**

Joshua Feldman conducted this project as a Keenan Research Summer Student at St. Michael's Hospital, University of Toronto. We thank Salvatore Vivoni, Nasheed Moqueet, and Ahmed Shakeri for helpful discussions. Sharmistha Mishra is supported by an Ontario HIV Treatment Network and Canadian Institute of Health Research New Investigator Award.

**FUNDING**

The study was supported by the Canadian Institutes of Health Research Foundation Grant FN 13455.

**STATEMENT OF COMPETING INTERESTS**

Declarations of interest: none